\documentclass[12pt,aps,preprint,superscriptaddress]{revtex4-2} 
\usepackage{graphicx,amsmath,amssymb,latexsym}
\usepackage[format=hang,hangindent=10pt]{subfig}
\newcommand{\eps}{\varepsilon}
\renewcommand{\vec}[1]{\mathbf{#1}}
\renewcommand{\Re}{\textrm{Re}\,}

\begin{document}
\title{Modeling of Subwavelength Gratings: Near-Field Behavior}

\author{Alexander Chernyavsky}
\affiliation{Institute of Automation and Electrometry,
Siberian Branch, Russian Academy of Sciences,
1 Koptyug Ave, 630090 Novosibirsk, Russia}
\affiliation{Novosibirsk State University,
2 Pirogov Str, 630090 Novosibirsk, Russia}
\author{Alexey Bereza}
\affiliation{Institute of Automation and Electrometry,
	Siberian Branch, Russian Academy of Sciences,
	1 Koptyug Ave, 630090 Novosibirsk, Russia}
\author{Leonid Frumin} 
\affiliation{Institute of Automation and Electrometry,
	Siberian Branch, Russian Academy of Sciences,
	1 Koptyug Ave, 630090 Novosibirsk, Russia}
\author{David Shapiro\footnote{email: \tt shapiro@iae.nsk.su}}
\affiliation{Institute of Automation and Electrometry,
	Siberian Branch, Russian Academy of Sciences,
	1 Koptyug Ave, 630090 Novosibirsk, Russia}
	
\date{\today}

\begin{abstract}Subwavelength gratings, with period shorter than the incident wavelength, have garnered significant attention in the fields of photonics, optoelectronics, and image sensor technology. In this research, we delve into the scattering characteristics of these gratings by employing the 2-dimensional point dipole approximation. Additionally, we propose a version of perturbation theory that relies on Fourier decomposition to obtain analytical expressions for the near-field behavior. We validate our models using numerical techniques such as boundary and finite element analysis. Notably, we explore how parameters like the grating period and slit width affect field enhancements. We demonstrate that our models produce qualitatively accurate results even for narrow slits.
\end{abstract}

\begin{titlepage}
\maketitle
\end{titlepage}
\section{Introduction}

Subwavelength gratings (SWGs), characterized by periods shorter than the wavelength of the radiation they interact with, have garnered growing interest and found significant applications across various fields. Their versatility has been demonstrated in enhancing the focusing of light beams \cite{cheng2014}, contributing to advances in silicon photonics \cite{wang2016}, enabling the development of optoelectronic devices exploiting surface plasmons  \cite{lee2017}, serving as mirrors in vertical-cavity surface-emitting lasers \cite{czyszanowski2017}, and playing a pivotal role in image sensor technology \cite{mateus2004,horie2017}. These structures provide compact and cost-effective solutions, even within the visible spectrum, offer exciting prospects for manipulating the polarization properties of emitted radiation \cite{Petrov2018}, and enhancing the performance of InGaAs infrared polarization imaging \cite{nano13182512}. The incorporation of double SWGs has shown promise in creating ultra-narrow resonances in the optical response of systems, with amplification factors of up to $10^5$ \cite{krylov2023}. Moreover, SWGs are very useful in hybrid multiplexing technology to improve optical communication capacity by combining multi-wavelengths, multi-modes, as well as dual-polarizations \cite{Yuan23}.

The electrodynamics theory of diffraction relies on Maxwell's equations and the decomposition of their solutions into eigenfunctions. This theory includes the Rigorous Coupled-Wave Analysis (RCWA), which has primarily been developed for configurations of high symmetry. Some examples of such configurations include relief profiles with square-wave, triangular, or saw-tooth shapes \cite{moharam1982}. Additionally, other notable cases involve  periodic arrays of parallel infinite circular cylinders \cite{twersky1962,kavaklioglu2002,belan2015}. In the case of relief profiles and ribbons, the eigenfunctions used are plane waves, whereas for cylindrical arrays, Bessel or Hankel functions serve as a basis. 

There are various numerical methods available for analyzing this problem. The current cutting-edge techniques include Finite-Difference Time-Domain (FDTD) \cite{yee1966}, Discrete Dipole Approximation (DDA) \cite{purcell73, Draine08}, Boundary Element Method (BEM) \cite{brebbia2016}, and Finite Element Method (FEM) \cite{zienkiewicz2005}. However, each of these methods comes with inherent limitations and some numerical errors that makes it difficult to compare these approaches. Consequently, there is a growing demand for new analytical solutions that can aid in optimizing photonic structures and validating numerical simulations.

Since the SWG period is much smaller than the radiation wavelength, a common simplification is to treat the grating as an effective continuous medium layer with averaged dielectric permittivity. This approach provides accurate results for reflection, transmission, and extinction when assuming that the effective medium is anisotropic \cite{nemykin2022}. However, this simple averaging method fails to capture intricate interference phenomena like resonances at the onset of new diffraction modes (Rayleigh frequencies) \cite{PhysRevB.74.245422,gomez2006} and near-field effects, like the sub-diffraction light focusing \cite{PRL08}. The goal of this study is to use analytical treatment of SWGs with cylindrical elements. We apply the 2D point dipole approximation, similar to the 3D model of the interaction between dielectric spheres by Markel \cite{Markel93}. Another approach used to get analytical expressions is the perturbation theory, valid for the dielectric permittivity  close to unity $|\eps-1|\ll1$. It differs from the quantum-mechanical Born series because of the specific boundary conditions in the electrodynamics. A rather similar approach was applied earlier to the scattering problem for two parallel cylinders \cite{bereza2017}.

Section \ref{s:dipoles} introduces the Point Dipole Approximation (PDA), enabling the derivation of simplified analytical formulas. Section \ref{s:perturbations} describes the perturbation series based on the Fourier series developed for the near-field calculations. Section \ref{s:numeric} is devoted to numerical modeling by  BEM and FEM to determine the near-field intensity card for electric and magnetic field, and comparing it to analytical formulas. Finally, Section \ref{s:conclusions} provides a summary of the conclusions.

\section{Point dipole approximation}\label{s:dipoles}

\subsection {Polarizability}
\begin{figure}\centering
\subfloat[\centering]{\includegraphics[width=0.47\textwidth]{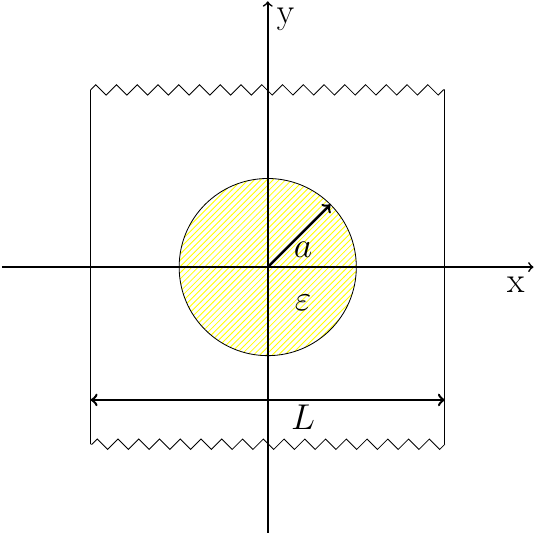}}%
\qquad
\subfloat[\centering]{\includegraphics[width=0.47\textwidth]{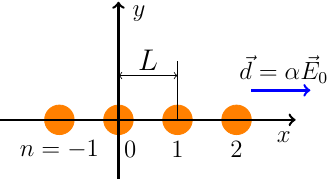}}		
\caption{(a) The cross-section of an elementary cell in the periodic grating with cylindrical elements. (b) The periodical chain of point dipoles denoted by circles.} \label{f:elemetary_cell}
\end{figure}

We are examining an infinite array of cylindrical structures with a radius of $a$ and a periodicity of $L$. An incident electromagnetic wave is coming from below along the $y$ axis in its positive direction, with our coordinate system oriented such that $z$ aligns with the axes of these parallel cylinders and $x$ is perpendicular to them. We are condensing our analysis to a fundamental unit cell, as depicted in Figure \ref{f:elemetary_cell}(a), while applying periodic boundary conditions at $x=-L/2$ and $x=L/2$. The circles  represent the cross-sections of the cylindrical elements, which have a dielectric permittivity of $\eps$. The surrounding medium is free space with $\eps=1$. The unique characteristics of the $p$-wave scattered by this grating when illuminated at normal incidence are primarily observed in the near field, specifically in the immediate vicinity of the cylinders. Below the investigation of subwavelength scattering within the lattice focuses on the study of the near-field behavior.

This approximation is highly justified for a sparsely distributed lattice, especially when the radius of a cylinder is much smaller than the lattice period. As depicted in Figure \ref{f:elemetary_cell}(b), in the case of a sparse lattice, the field affecting each cylinder is approximately uniform. The Point Dipole Approximation (PDA) can also be applied, to some extent, for a densely packed lattice when the gap dimension $\delta$ between neighboring cylinders is relatively small, i.e., when $\delta=L-2a\ll L$. However, a noticeable error arises when assessing the interaction of a "dimer," which is a pair of adjacent cylinders, because the field near a polarized cylinder deviates from a simple dipole-like behavior. In such cases, it becomes necessary to consider additional terms in the multipole expansion.

The polarizability of a cylinder aligns well with the PDA and ultimately results in an effective lattice polarizability coefficient that differs from that of an individual cylinder. At normal incidence, the phase of incident field is the same for all cylinders. Consequently, the two-dimensional dipoles excited in them have the same phase. When evaluating the near-field effects from these dipoles in the subwavelength regime, our calculations can be treated in the electrostatic limit, disregarding variations in the phase of the individual dipoles at the observation point.

The electric field vector of the incident wave lies along $x$-axis
\[
\vec E_x=E_0\vec e_x,
\]
where $\vec e_x$ is the unit orth. Total field 
$\vec E$ acting on each cylinder is the sum of external field $\vec E_0$ and the field of all neighboring dipoles 
$\vec E'$:
\[
\vec E=\vec E_0+\vec E'.
\]
The dipole moment per unit length $\vec d$ is proportional to the total field $\vec E$: $\vec d=\alpha\vec E,$ where $\alpha$ is the polarizability of dielectric cylinder in a homogeneous external field \cite{smythe1950}:
\begin{equation}\label{polarizability}
	\alpha=\frac{a^2}2\frac{\eps-1}{\eps+1}.
\end{equation}
Here $\eps$ is the dielectric constant of the cylinder material.

A two-dimensional dipole with number $n$ directed along the $x$-axis creates field
\[
\vec E_n'=\frac{2\vec d}{(nL)^2},\quad n\neq0
\]
in the neighborhood of ``central'' cylinder with number 0.
In the sub-wavelength (electrostatic) limit, the total field from all other cylinders at the center of cylinder $n=0$ ($x=0$) is given by the sum:
\[
\vec E'=2\sum_{n=1}^\infty \vec E_n'=\frac{2\pi^2}{3L^2}
\alpha(\vec E_0+\vec E').
\]
Hence we obtain for the field of other dipoles and the dipole moment induced:
\begin{align*}
\vec E'=\frac{2\pi^2\alpha}{3L^2-2\pi^2\alpha}\vec E_0,
\label{dipole}\\
	\vec d=\alpha(\vec E_0+\vec E')=\frac{3L^2\alpha}{3L^2-2\pi^2\alpha}\vec E_0.
\end{align*}
Using formula (\ref{polarizability}), we get
\begin{equation}\label{dipole-transverse}
	\vec d= \alpha_{g}\vec E_0.
\end{equation}	
In this context, we can refer to the proportionality factor 
\[
\alpha_{g}=\frac{3L^2\alpha}{3L^2-2\pi^2\alpha}=\frac{3 a^2L^2(\eps-1)}{6L^2(\eps+1)-2a^2\pi^2(\eps-1)}
\]
as the ``grating polarizability'' Eq. (\ref{dipole-transverse}) highlights that the resonant field enhancement in the slits occurs in regions where the denominator is small. This phenomenon corresponds to the lattice dipole plasmon resonance. Specifically, such a resonance can be observed in a lattice composed of metallic cylinders in the optical frequency range where the real part of the dielectric constant, $\Re\eps$, is less than zero.

To achieve a strong resonance, it is essential to consider a "heterogeneous" lattice configuration with narrow slits between the cylinders. Unlike a dimer, which is not accurately described in PDA, the lattice resonance is a collective effect involving all the cylinders. For closely spaced scatterers, PDA still introduces some error, although it is not as pronounced as it is for a dimer configuration. 

\subsection{Near field}
In the near-field approximation the field of $n$-th dipole at the observation point with Cartesian coordinates $x,y$ is described by the expression
\[
\vec E_n=\frac{4\vec r_n(\vec r_n\cdot\vec d)}{r_n^4}-
\frac{2\vec d}{r_n^2},
\]
where the dot inside parentheses denotes the scalar product, $\vec r_n=\vec e_x x+\vec e_y y$, vectors $\vec e_{x,y}$ are unit orthants of the coordinate axes, $x_n=x-nL$, $r_n=\sqrt{x_n^2+y^2}$. 

The total field $E_x$ from all cylinders
of the lattice, directed along the $x$-axis, is described by the infinite sum:
\[
E_x=\sum_{n=-\infty}^\infty
(\vec e_x\cdot \vec E_n)=d\sum_{n=-\infty}^\infty
\frac{-y^2+(x-Ln)^2}{[y^2+(x-Ln)^2]^2}.
\]
For the $y$-component of the cylinder field $E_y$ at the observation point, respectively, we obtain:
\[
E_y=\sum_{n=-\infty}^\infty
(\vec e_y\cdot \vec E_n)=d\sum_{n=-\infty}^\infty
\frac{4y(x-Ln)}{[y^2+(x-Ln)^2]^2}.
\]
Note that both sums, give the periodic function of $x$.

Summing up the series, we get:
\begin{align}\label{Ex}
	E_x=\frac{\pi^2}{L^2}d\left[\csc(\pi(x-iy)/L)+
	\csc(\pi(x+iy)/L)\right],\\
	\label{Ey}
	E_y=\frac{i\pi^2}{L^2}d\left[\csc(\pi(x-iy)/L)-
	\csc(\pi(x+iy)/L)\right].
\end{align}
The total electric field intensity of the scattered
radiation is given by a simple formula:
\begin{equation}\label{intensity}
	I_s=|E_x|^2+|E_y|^2=\frac{16\pi^4d^2}{L^4\left(\cos\frac{2\pi x}L-\cosh\frac{2\pi y}L\right)^2}.
\end{equation}
From Eq. (\ref{intensity}), we can observe that the intensity of the scattered radiation exhibits periodic variations along the $x$-axis. In contrast, along the $y$-axis, the intensity diminishes symmetrically and quickly. 

\begin{figure}\centering
	\includegraphics[width=5in]{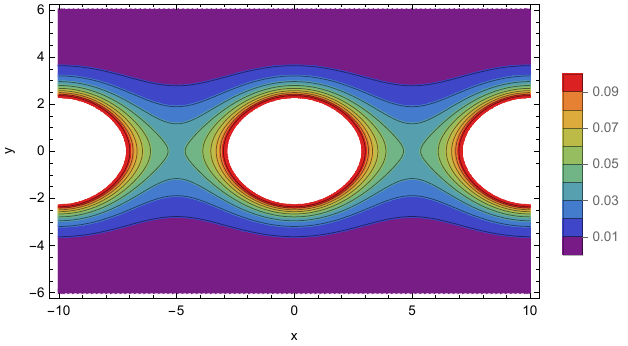}
	\caption{The electric field intensity distribution $|\vec E|^2$ of the scattered radiation of the grating modeled by PDA (\ref{intensity}) at $L=10$ (in $d^2/L^4$ units).  In this representation, the white ovals highlight areas excluded from the diagram since their intensity exceeds 0.1.}\label{f:contours}
\end{figure}

Figure \ref{f:contours} displays the distribution of scattered field intensity near a cylinder, in PDA. One can see saddle points near the center of a gap. The positions of saddle points are $x=(n+1/2)L$ and $y=0$, according to Eq. (\ref{intensity}). It's important to note that this model results in an overestimation of the field intensity near the point dipole itself. Due to this overestimation, we have not included the regions around points $x=nL, n=0,\pm1,\dots,$ and $y=0$ in the figure. 

\section{Perturbation theory}\label{s:perturbations}
\subsection{Fourier Series}
PDA model provides a reasonably accurate description of the scattered electric field in the outer region, outside the cylinders. However, inside each cylinder, the model predicts a singularity, which is a point where the field becomes infinite. Additionally, this model does not account for the scattered magnetic field, as it effectively sets the rotor (curl) of the magnetic field to zero. Nevertheless, in reality, the scattered magnetic field is not zero. To address these limitations and obtain a more accurate representation of the electric field, we propose to consider the following equation: 
\begin{equation}\label{Maxwell}
	\nabla\times\nabla\times\vec E=ik_0\nabla\times\vec H=\eps(\vec r)k_0^2\vec E.
\end{equation}
Derived from Maxwell's equations, it describes field $\vec E$ in a medium with magnetic vacuum permeability $\mu$=1 and with space-variable dielectric permittivity $\eps(\vec r)$. 

The scattering of $p$-wave is commonly described using a single Helmholtz equation for the magnetic field. However, solving a periodic problem for a magnetic field using the Fourier method can be challenging due to the discontinuity of its derivative at the boundary caused by the imposed boundary conditions. To address this issue and effectively solve the problem, a pair of equations for a two-dimensional electric field is employed below. These equations ensure that the electric field is not only continuous but also maintains continuity with its derivative, thus facilitating a more tractable solution.

In the case of equidistant parallel cylinders, function $\eps(\vec r)$ is expressed in terms of the theta-functions:
\begin{equation}
\eps(\vec r)=1+(\eps-1)\sum\limits_{m=-\infty}^\infty
\theta\left(a-\sqrt{(x-Lm)^2+y^2}\right).
\end{equation}
For $p$-wave $E_z=0$, and components $E_x(x,y)$ and $E_y(x,y)$ remain in the equation (\ref{Maxwell}). Represent the components, the periodic functions of $x$, as Fourier series:
\begin{equation}\label{electric_field}
	E_x=\sum\limits_{m=-\infty}^\infty
	U_m(y)e^{2\pi m x/L},\quad E_y=\sum\limits_{m=-\infty}^\infty
	V_m(y)e^{2\pi m x/L}.
\end{equation}
At normal incidence of the external field this representation is natural. Substituting the sums, multiplying by $\exp(-2\pi i px/L)$, and integrating over an elementary cell, Fig. \ref{f:elemetary_cell}(a), from $-L/2$ to $L/2$, we obtain the following equations for the amplitudes:
\begin{align}
U_p''-\frac{2\pi i p}{L}V_p'+k_0^2U_p\nonumber\\=-(\eps-1)\theta(a-|y|)
\sum\limits_{m=-\infty}^\infty\frac{U_m(y)}{\pi(m-p)}
\sin\frac{2\pi(m-p)\sqrt{a^2-y^2}}{L},\nonumber\\
-\frac{4\pi^2p^2}{L^2}V_p-\frac{2\pi i p}{L}U_p'+k_0^2V_p\nonumber\\=-(\eps-1)\theta(a-|y|)
\sum\limits_{m=-\infty}^\infty\frac{V_m(y)}{\pi(m-p)}
\sin\frac{2\pi(m-p)\sqrt{a^2-y^2}}{L}.\label{modes}
\end{align}

Now express the equations for first-order amplitudes, with more clarity. Starting from Eq. (\ref{modes}), we can write these equations as follows: $U_p=U^{(0)}_p+U^{(1)}_p+\dots, V_p=V^{(0)}_p+V^{(1)}_p+\dots$
Let us consider the right side in Eq. (\ref{modes}) as small and write the amplitudes as perturbation series: $U_p=U^{(0)}_p+U^{(1)}_p+\dots, V_p=V^{(0)}_p+V^{(1)}_p+\dots$ 
For the unperturbed, zero-order solution, we have: $U_{p\neq0}^{(0)}=V_{p\neq0}^{(0)}=0$; otherwise  exponential increasing along $y$ would arise. $V_{0}^{(0)}=0$, due to the perpendicular incidence. We will take for definiteness $U_{0}^{(0)}=e^{ik_0y}$, thus the electric field at zero order has the specific physical meaning of the external field falling from bottom to top onto the grating. Equations for the first-order amplitudes $U_{p}^{(1)}, V_{p}^{(1)}$  are
\begin{align}
U_p^{(1)\prime\prime}-\frac{2\pi i p}{L}V_p^{(1)\prime}+k_0^2U_p^{(1)}=-(\eps-1)\theta(a-|y|)
\frac{e^{ik_0y}}{\pi p}
\sin\frac{2\pi p\sqrt{a^2-y^2}}{L},\nonumber\\	
q_p^2V_p^{(1)}+\frac{2\pi i p}{L}U_p^{(1)\prime}=0,\quad q_p^2=\frac{4\pi^2p^2}{L^2}-k_0^2.
\end{align}	

\subsection{Solutions}
At $p\neq0$ we find a solution decreasing at infinity:
\begin{align}
U_p^{(1)}=-(\eps-1)q_p^2\int_{-a}^a G_p(y,\Tilde{y})\frac{e^{ik_0\Tilde{y}}}{\pi p}\sin\frac{2\pi p\sqrt{a^2-\Tilde{y}^2}}{L}\,d\Tilde{y},\nonumber\\	
V_p^{(1)}=-\frac{2\pi i p}{Lq_p^2}U_p^{(1)\prime},\quad
G_p(y,\Tilde{y})=-\frac{e^{-q_p|y-\Tilde{y}|}}{2q_p}.\label{Green-function}
\end{align}
At $p=0$ there is no decreasing, then use different approach. First, we take any Green's function and describe the general solution
\begin{align}
U_0^{(1)}=C_0^{(1)}e^{ik_0y}+D_0^{(1)}e^{-ik_0y}\nonumber\\
-(\eps-1)k_0^2\int_{-a}^a
\left[\theta(y-\Tilde{y})\frac{e^{ik_0(y-\Tilde{y})}
-e^{-ik_0(y-\Tilde{y})}}{2ik_0}\right]
\frac{2\sqrt{a^2-\Tilde{y}^2}}{L}e^{ik_0\Tilde{y}}
\,d\Tilde{y}.\label{first-order}
\end{align}
The convenient Green's function to be the one that is zero when $y<\Tilde{y}$.

The coefficients $C_0^{(1)}$ and $D_0^{(1)}$ are calculated from the scattering boundary conditions:
\begin{equation}
U_0(y\to-\infty)=e^{ik_0y}+A_0e^{-ik_0y},\quad 
V_0(y\to+\infty)=B_0e^{ik_0y}.\label{Jost}
\end{equation}
Substituting $U_0=U_0^{(1)}+U_0^{(1)}$ into condition (\ref{Jost}) we get
\begin{equation}\label{first-order-coefficients}
C_0^{(1)}=0,\quad D_0^{(1)}=A_0^{(1)}=\frac{i(\eps-1)k_0}{2L}
\pi a^2\left[J_0(2k_0a)+J_2(2k_0a)\right],
\end{equation}
where $J_0,J_2$ are the Bessel functions of the zeroth and second order, respectively \cite{Olver10}. Within the first order $B_0^{(1)}=i(\eps-1)k_0\pi a^2/2L$, i.e. practically coincide with $A_0^{(1)}$. At last, the magnetic field can be found as a rotor of electric field vector:
\begin{equation}\label{magnetic_fiels}
	H_z(x,y)=\frac1{ik_0}\left[(E_y)_x-(E_x)_y\right]
	=-\frac{U_0^{(1)\prime}(y)}{ik_0}-ik_0\sum\limits_{p\neq0}
	\frac{U_p^{(1)\prime}(y)}{q_p^2}e^{2\pi p x/L}.
\end{equation}
\begin{figure}\centering
\subfloat[\centering]{\includegraphics[width=0.47\textwidth]{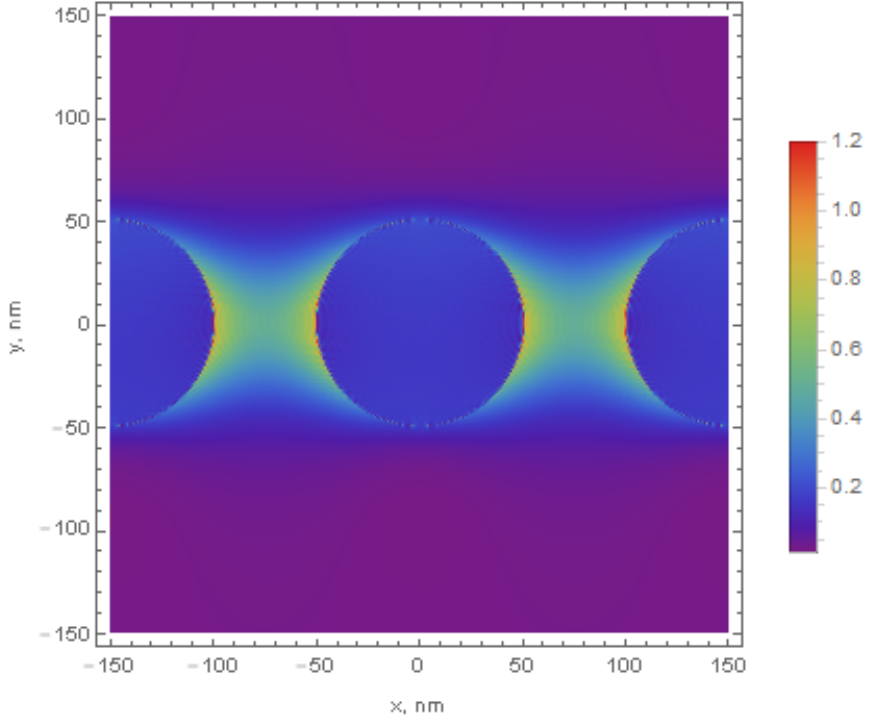}}%
\qquad
\subfloat[\centering]{\includegraphics[width=0.47\textwidth]{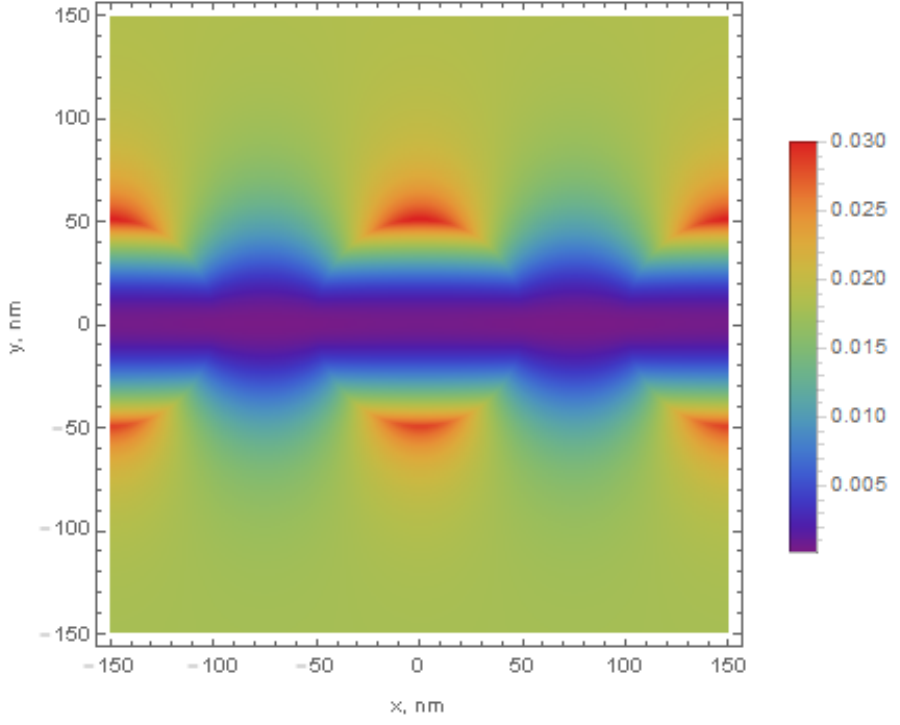}}		
\caption{Intensity distributions near the grating at $\eps=2.25, \lambda=1.512~\mu$m, $a=50~$nm, $L=150~$nm within the first order perturbation theory: (a) $|\vec E|^2$ calculated by Eq. (\ref{electric_field}),  (b) $|\vec H|^2$ from Eq. (\ref{magnetic_fiels}).}\label{f:perturbations}
\end{figure}

Estimating from Eq. (\ref{first-order-coefficients}), we can determine the small perturbation parameter by comparing the first and zeroth orders, given as:
\begin{equation}
	|\eps - 1|(k_0a)\frac{a}{L} \ll 1.
\end{equation}
This observation suggests that the method remains valid for SWG, even when $|\eps - 1| \sim 1$, while $k_0a\ll1$. The  expansion is particularly effective for sparse gratings when $a\ll L$.

The first-order correction is illustrated in Fig. \ref{f:perturbations}, which includes 221 harmonics (with $|m|\leqslant 110$) chosen based on their ability to minimize phase variations between adjacent cylinders ($\delta\varphi\ll 1$). In these plots, you can observe the maximum values of the scattered electric field intensity at the interfaces between the dielectric and free space, both on the right and left, see subfigure (a). There are saddle points at the geometric centers of the slits, similar to what one see in the PDA. The distribution of magnetic field shows its maximal value at the top and bottom of the cylinders. Another distinctive feature of the magnetic field is the region of low field intensity at $|y|< 0.25~\mu$m in subfigure (b). This minimum results from strong reflection from the SWG.

Consequently, employing the perturbation theory enables us to eliminate the singularity near the center of a cylinder, which is a characteristic feature of the PDA. As a result, the first-order perturbation theory provides a deeper understanding of the distributions of electric and magnetic field intensities. The analytical expressions derived from this approach have significant potential for optimizing resonant periodic structures.

\section{Numerical modeling}\label{s:numeric}
\begin{figure}\centering
\subfloat[\centering]{\includegraphics[width=5in]{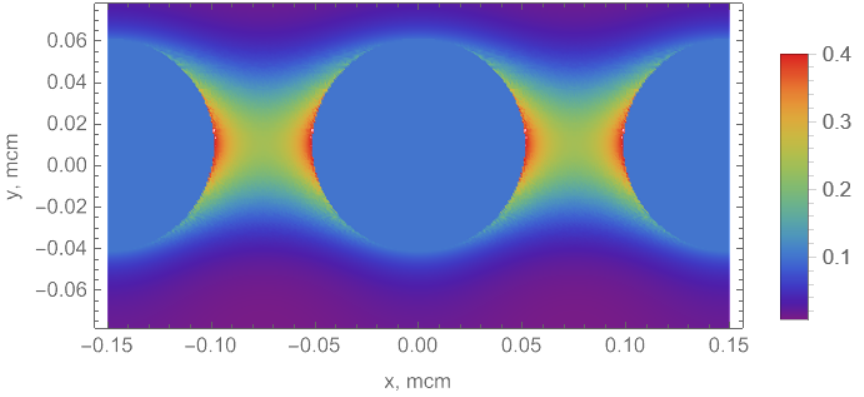}}

\subfloat[\centering]{\includegraphics[width=5in]{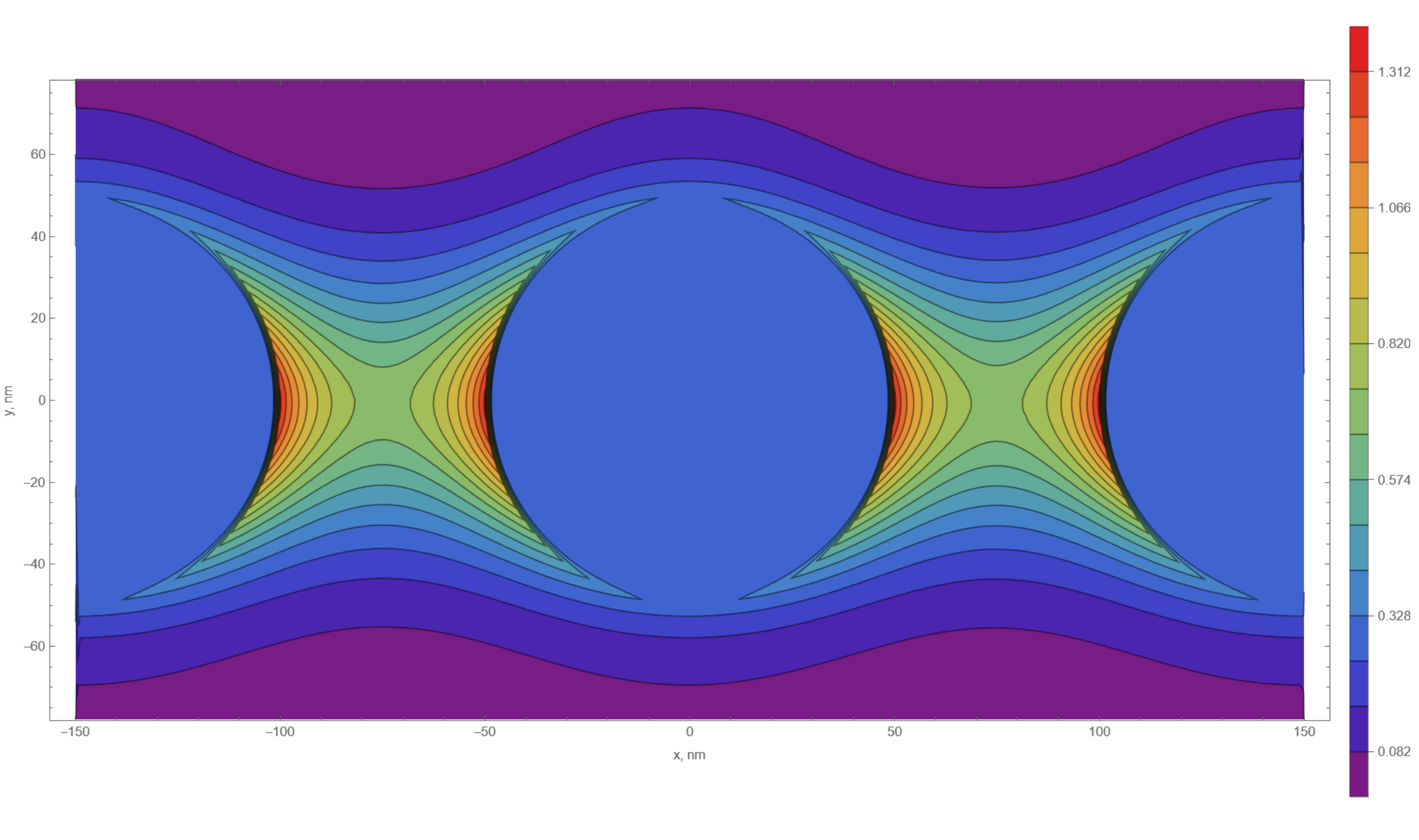}}
\caption{Distributions near the grating at $\eps=2.25, \lambda=1.512~\mu$m, $a=50~$nm, $L=150~$nm: $|\vec E|^2$   calculated by BEM (a),  FEM (b).}\label{f:BEM}	
\end{figure}

To determine the scattered field, at first we utilized BEM, leveraging Floquet's theorem. Our approach involved computing an effective Green's function represented as a series. The theorem simplified problem by focusing on the scattering by a single unit cell. Unlike the previous paper (documented in \cite{periodic2013}), our current calculation does not consider the influence of a dielectric substrate and then becomes simpler. For visual representation, please refer to Fig. \ref{f:BEM}(a), which displays the intensity distribution in the near-field domain out of the cylinders. Inside a cylinder the field is not calculated. The result obtained for a quite sparse lattice by FEM, in COMSOL Multiphysics\textsuperscript{\textregistered} \cite{COMSOL22}, is shown in Fig. \ref{f:BEM}(b) as a contour plot.  Here we treat a finite grating with $N=20$ cylinders while only central part is shown. 

\begin{figure}\centering
	\includegraphics[width=5in]{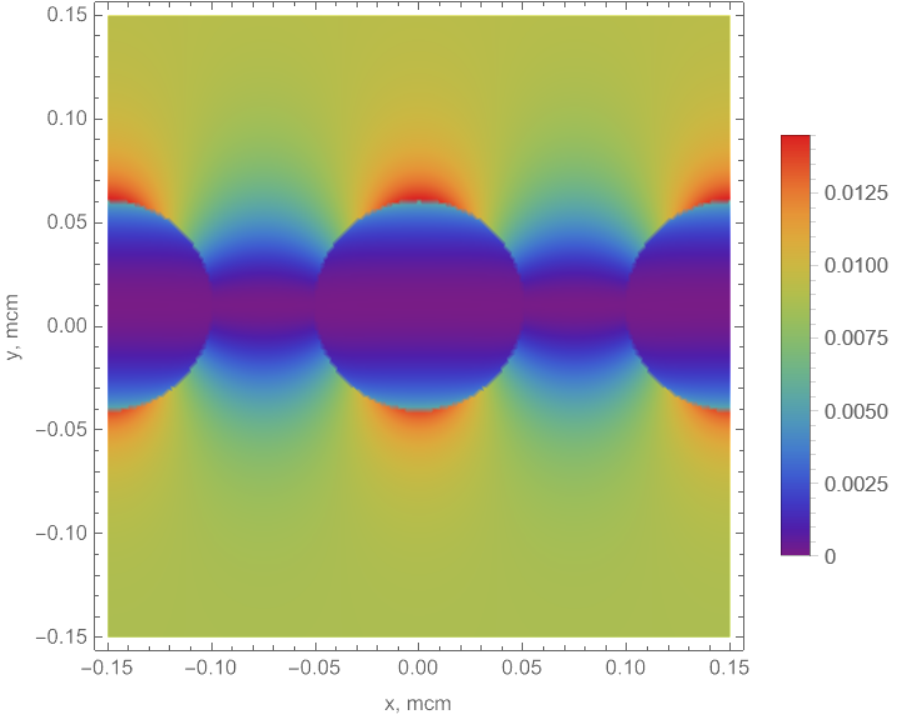}
	\caption{Distributions near the grating at $\eps=2.25, \lambda=1.512~\mu$m, $a=50~$nm, $L=150~$nm:  $|\vec H|^2$ calculated by BEM.}\label{f:BEM_H}	
\end{figure}

When we compare Fig. \ref{f:BEM}(a,b) to PDA depicted in Fig. \ref{f:contours}, we observe similar patterns. In both BEM and FEM calculations, saddle points are evident at the midpoint between neighboring cylinders, while maxima are attained near the interface between free space and the dielectric medium. 
Figure \ref{f:BEM_H} shows the squared magnetic field computed using BEM. The magnetic field doesn't require finding derivatives, so the computational domain encompasses the entire diagram, including internal areas. The distribution of magnetic field reaches its peak near the dielectric material at the top and bottom points of the diagram. 
 
When comparing the patterns in Figs. \ref{f:BEM}, \ref{f:BEM_H} to the representations provided by the perturbation theory in Fig. \ref{f:perturbations}, it becomes evident that there are striking similarities between the two sets of results. These similarities include the presence of maxima in the electric field at points on the right and left, maxima in the magnetic field at the top and bottom, the existence of saddle points, and the appearance of a minimum strip at small values of $y$. 

\begin{figure}\centering
\includegraphics[width=5in]{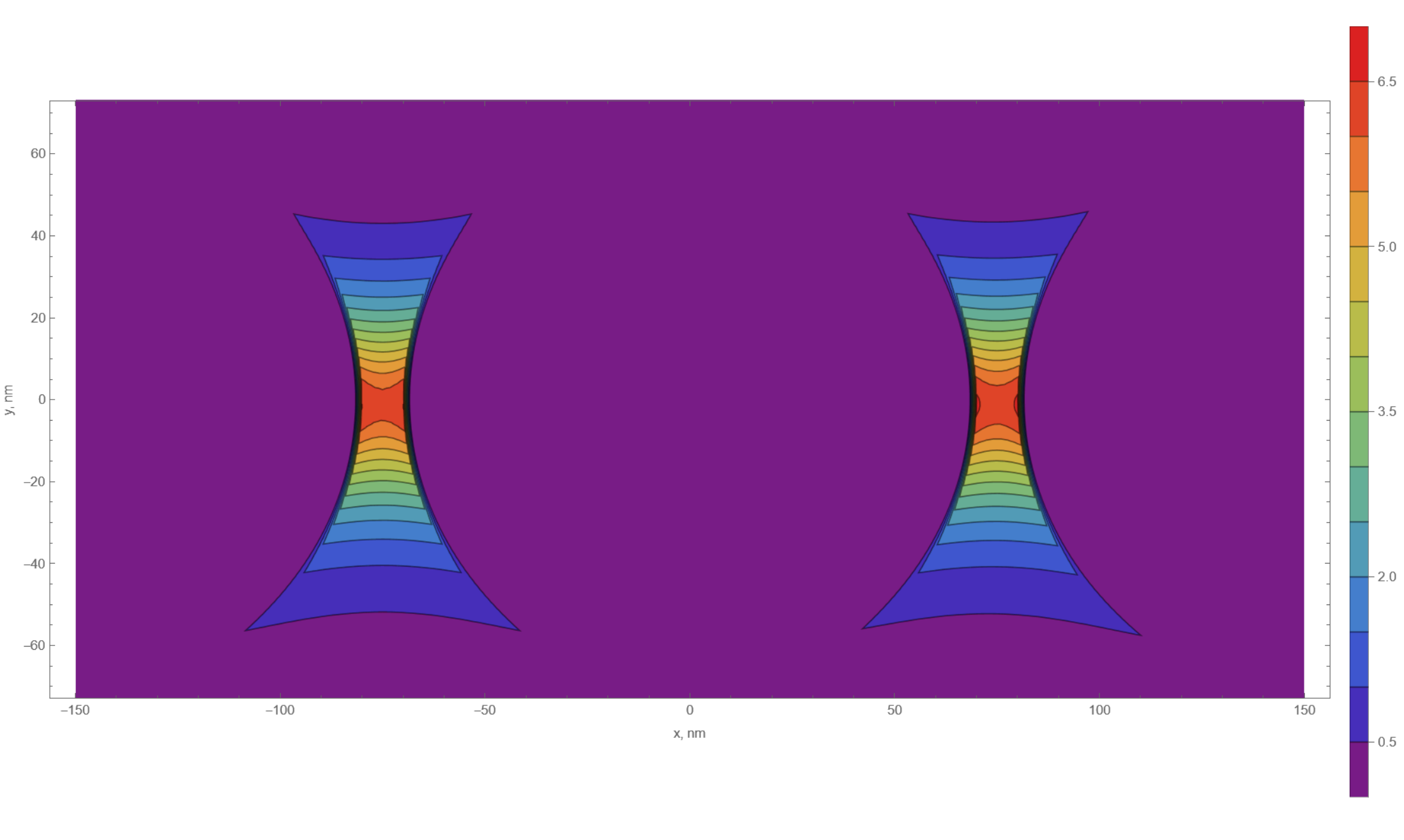}
\caption{Contour plot of intensity distribution, calculated by  FEM, at the same parameters as in Fig. \ref{f:BEM}, but for greater radius $a=70~$nm.}\label{f:narrow}	
\end{figure}
We repeat calculations for the same period $L$, but with a narrow slit $\delta=10$~ nm. FEM yields the saddle points, too. The field enhancement factor increases by approximately 6 times in this case, as shown in Fig. \ref{f:narrow}. 

\section{Conclusions}\label{s:conclusions}
 
Subwavelength gratings have a wide range of applications in photonics and optoelectronics. In this study, we analyze a periodic system of parallel cylinders. We use the 2-dimensional point-dipole approximation and Born series. We have derived analytical expressions for the electric and magnetic components in the near field. To validate our analytical model, we performed numerical simulations that prove its accuracy. Remarkably, we identified the emergence of saddle points at the geometric centers of the slits. Numerical simulations confirmed this observation. It is shown that the dipole approximation provides the correct qualitative description of the behavior in the slit, while the perturbation theory offers an approximate quantitative representation over the entire $(x, y)$ plane. The findings from this research  will enhance understanding facilitates optimized designs and pave the way for their application in cutting-edge optical technologies.

\section{Acknowldgements}
The authors are grateful to O.V. Belai,  A.V. Nemykin, and S.V. Perminov for helpful discussions. The work was funded by Russian Science Foundation, grant \#22-22-00633.

%

\end{document}